\documentclass{emulateapj}  

\usepackage{graphicx}  
\usepackage{apjfonts}

\slugcomment{Accepted for publication in ApJ, 2009 July 20 v700 issue}

\shorttitle{Supersonic horizontal 
flows in quiet Sun granules} 
\shortauthors{L.R. Bellot Rubio}
\begin{document}

\title{Detection of supersonic horizontal flows in the solar granulation} 

\author{L.R. Bellot Rubio}

\affil{Instituto de Astrof\'{\i}sica de
Andaluc\'{\i}a (CSIC), Apdo. de Correos 3004, 18080 Granada, Spain} 
\email{lbellot@iaa.es}

\begin{abstract}
Hydrodynamic simulations of granular convection predict the existence
of supersonic flows covering $\sim$3-4\% of the solar surface at any
time, but these flows have not been detected unambigously as yet.
Using data from the spectropolarimeter aboard the Hinode satellite, I
present direct evidence of fast horizontal plasma motions in quiet Sun
granules. Their visibility increases toward the limb due to more
favorable viewing conditions. At the resolution of Hinode, the
horizontal flows give rise to asymmetric intensity profiles with very
inclined blue wings and even line satellites located blueward of the
main absorption feature. Doppler shifts of up to 9~km~s$^{-1}$ are
observed at the edges of bright granules, demonstrating that the flows
reach supersonic speeds. The strongest velocities occur in patches of
0.5\arcsec\/ or less. They tend to be associated with enhanced
continuum intensities, line widths, and equivalent widths, but large
values of these parameters do not necessarily imply the existence of
supersonic flows. Time series of spectropolarimetric measurements in
regions away from disk center show the transient nature of the strong
horizontal motions, which last only for a fraction of the granule
lifetime. Supersonic flows are expected to produce shocks at the
boundaries between granules and intergranular lanes, and may also play
a role in the emergence of small-scale magnetic fields in quiet Sun
internetwork regions.

\end{abstract}


\section{Introduction}
\label{sec:intro}
Granules and intergranular lanes are the manifestation of overshooting
convection in the solar photosphere, a process that has been studied
in detail by means of 2D and 3D simulations (e.g., Stein \& Nordlund
1989; Cattaneo et al.\ 1990; Steffen \& Freytag 1991; Stein
\& Nordlund 1998; Steiner et al.\ 1998; Gadun et al.\ 1999; Ploner et
al.\ 1999). One of the most striking results of the simulations is the
existence of supersonic horizontal flows in granules. As explained by
Stein \& Nordlund (1998), these flows are a natural consequence of mass
conservation in a highly stratified atmosphere. The vertical upflows
of granules turn into horizontal flows relatively soon due to the
exponential decrease of the density with height. All the mass emerging
into the photosphere through the granular cross section must leave the
granule through its edges. Since the lateral area available for the
outflow is much smaller than the cross section of the granule, the
horizontal flow is forced to accelerate to conserve mass. The larger
the granule, the more pronounced the area difference and the stronger
the horizontal flow. Occasionally, supersonic velocities are reached.

The detection of these motions is challenging for a number of
reasons. First of all, they require spectroscopic measurements near
the limb, in order to maximize the projection of the horizontal
velocity to the line of sight. Also, high angular resolution is needed
to identify the spectral signatures of strong flows, which may
disappear if atmospheres with different properties are mixed in the
resolution element. Unfortunately, the spatial resolution is greatly
reduced towards the limb because of geometrical foreshortening and
because of the lower intensity contrast, which decreases the
performance of adaptive optics systems.

Given these difficulties, the efforts have concentrated
on the detection of shocks rather than the supersonic flows
themselves. According to the simulations, shocks can be produced by
the deceleration of transonic flows in intergranular lanes or by the
collision of fast flows from adjacent granules (Cattaneo et al.\ 1990;
Stein \& Nordlund 1998), with signatures that are visible on larger
scales. This means that the requirement of extremely high spatial
resolution can be relaxed to some degree. In addition, shock
observations are possible closer to the disk center (because the
detection does not rely on Doppler shift measurements).

The first indications of shocks were obtained by Nesis et al.\ (1992),
who observed enhanced line widths in intergranular lanes and
interpreted them as increased turbulence due to the passage of shock
fronts. Solanki et al.\ (1996) also detected broad intensity profiles
compatible with those expected from shocks. More recently, Ryb\'ak et
al.\ (2004) have identified clear shock signatures in spectroscopic
observations of a granular region taken with adaptive optics. For the
first time, the temporal evolution of a shock event could be followed
and compared with predictions from numerical simulations.

While the existence of shocks can be considered well established by
now, the motions causing them have not been observed yet. The purpose
of this paper is to provide unequivocal evidence of supersonic
horizontal flows in solar granular convection. They are detected as
strongly Doppler shifted satellites in the \ion{Fe}{1} 630.2~nm lines
recorded by Hinode away from disk center. Their signatures and basic
properties are studied in \S 3. \S 4 and \S 5 deal with the spatial
distribution and temporal evolution of the flows. Finally, \S 6
summarizes the main results of this work.

\begin{figure*}[t]
\resizebox{.309\hsize}{!}{\includegraphics[bb=43 360 327 813]{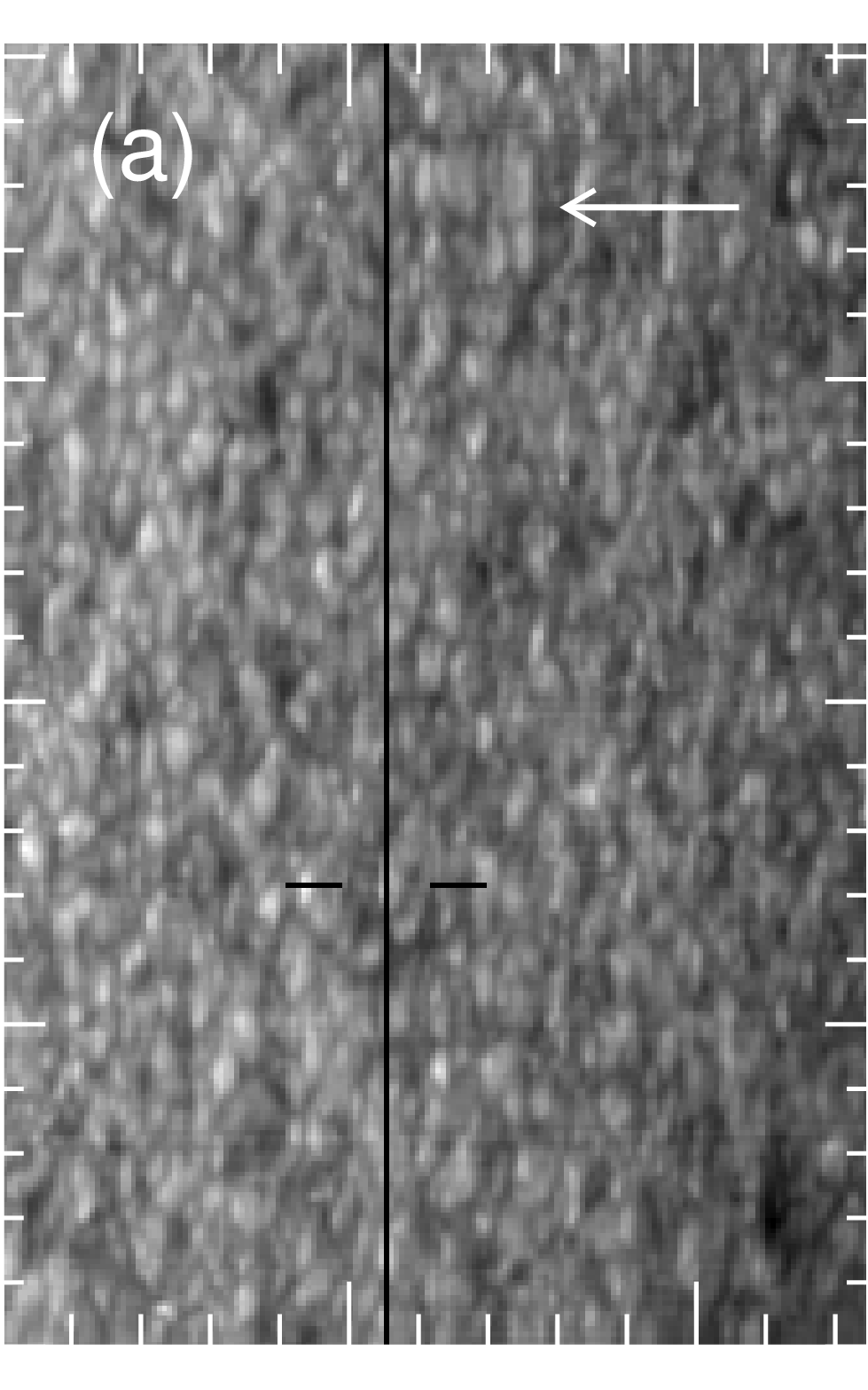}}
\resizebox{.309\hsize}{!}{\includegraphics[bb=85 360 369 816]{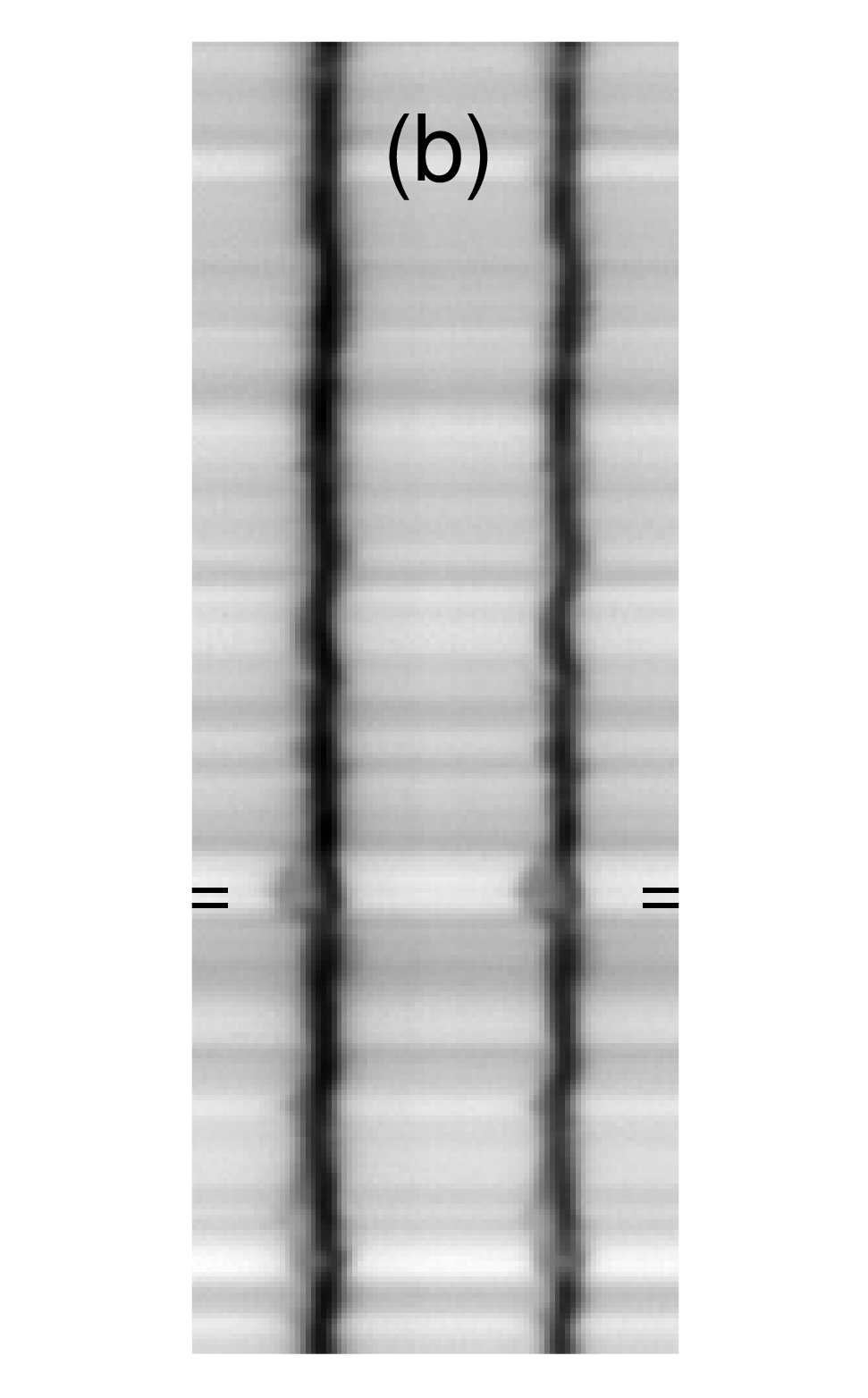} }
\resizebox{.375\hsize}{!}{\includegraphics[bb=134 363 469 816]{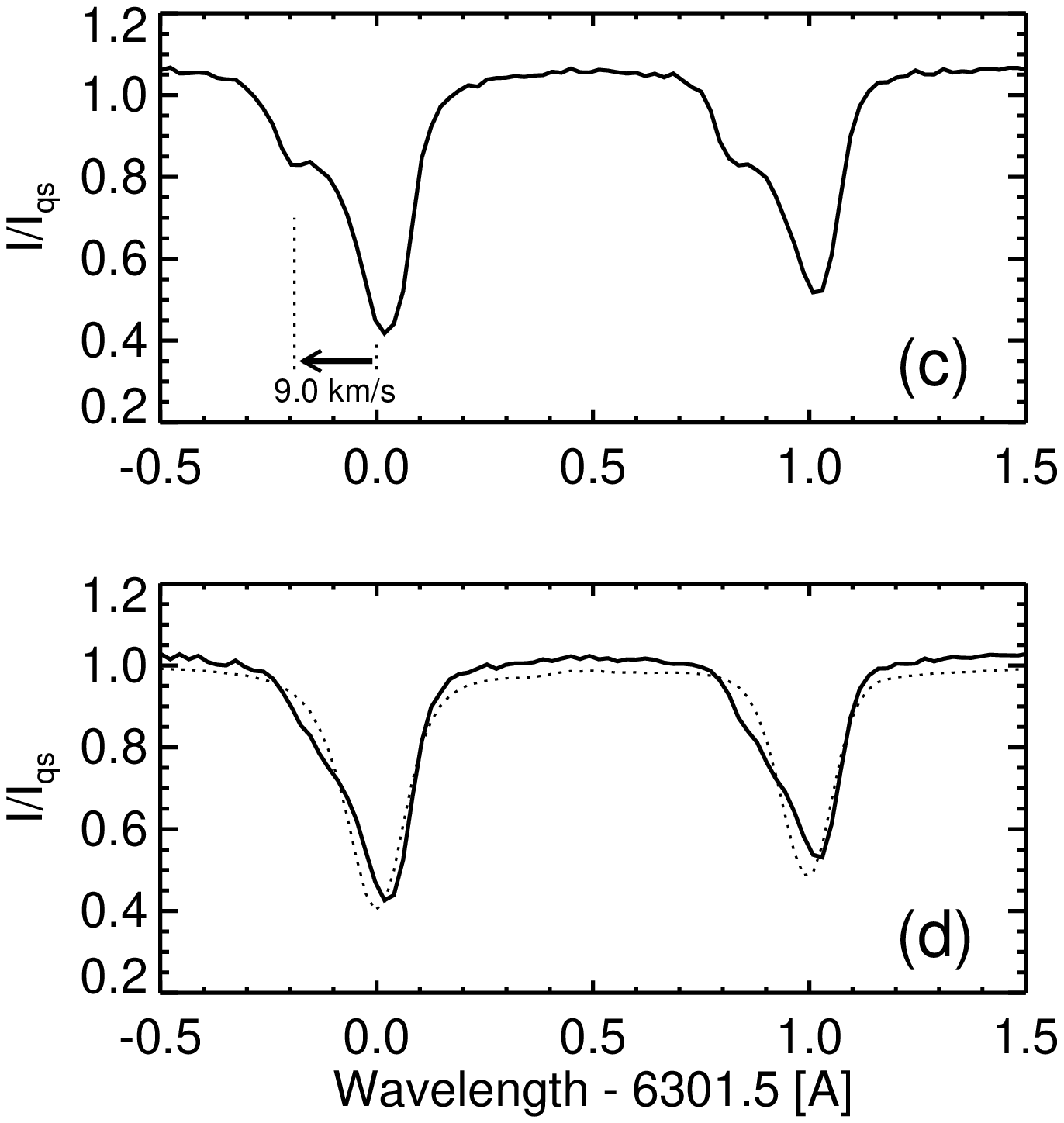}} 

\caption{{\em (a)} Quiet Sun region observed with the Hinode/SP on 
September 9, 2007. The figure displays a field of view (FOV) of
$24.8\arcsec \times 40.5\arcsec$, with North at the top and West to
the right. Tickmarks are separated by 2\arcsec. The heliocentric
distance varies between 70$^\circ$ and 75$^\circ$ ($\mu=0.34$ and
0.26) from left to right. The arrow indicates the direction to disk
center. The vertical line marks a slit position that crosses several
granules, one of which is identified with horizontal dashes. {\em (b)}
Intensity profiles of the
\ion{Fe}{1}~630.15 and 630.25~nm lines along the slit displayed in
panel {\em a}. Wavelength increases from left to right, covering an
interval of 0.2~nm.   Two spatial positions are
selected for further analysis (short horizontal dashes). The distance
between them is 0.48\arcsec. {\em (c)} Intensity profiles emerging from 
the upper spatial position indicated in panel {\em b}. The zero of the
wavelength scale corresponds to the line core position of the average
\ion{Fe}{1} 630.15 nm profile in the quiet Sun. {\em (d)} Same, for the
other spatial position marked in panel {\em b}. The dotted line represents
the average quiet-Sun profile.}
\label{fig1}
\end{figure*}

\section{Observations and data analysis}
\label{sec:obs}
The observations analyzed here were taken with the spectropolarimeter
(SP; Lites et al. 2001) of the Solar Optical Telescope (Tsuneta et
al.\ 2008) aboard the Hinode satellite (Kosugi et al.\ 2007). This
instrument measures the polarization spectra of the
\ion{Fe}{1} lines at 630.15 and 630.25~nm. The wavelength sampling is
2.153~pm~pixel$^{-1}$. With a slit width and a pixel size of
0.16\arcsec, the Hinode/SP achieves a spatial resolution of about
0.32\arcsec. 

The data have been corrected for dark current, flatfield, and various
instrumental effects using the sp\_prep.pro routine included in the
SolarSoft package. Three quiet Sun regions observed on 2007 September
9, 15, and 24 at heliocentric angles between 40$^\circ$ and 75$^\circ$
will be considered in this paper. The measurements were taken in the
framework of the {\em Hinode Operation Plans} 14 and 25, both of which
involved joint campaigns with the four solar telescopes of the Canary
Islands (Spain). The observed \ion{Fe}{1} 630.15~nm profiles have been
used to compute continuum intensities, residual intensities,
equivalent widths, full widths at half maximum, line-core velocities,
and line bisectors at different intensity levels. In addition, the
total polarization has been determined as ${\rm TP} =
\int (Q^2+U^2+V^2)^{1/2} /I_{\rm qs} \, {\rm d}\lambda$, where 
$I_{\rm qs}$ is the quiet Sun continuum at 630.2~nm. The integral
extends over a wavelength range of 0.1098~nm to encompass the
\ion{Fe}{1}~630.15~nm line.

\section{Signatures of supersonic horizontal flows}

At high spatial resolution, supersonic granular flows leave distinct
spectroscopic signatures in the form of line satellites and very
asymmetric intensity profiles. Examples of such fingerprints are
given in this Section.

Figure 1{\em a} shows the continuum intensity map of a quiet Sun region
observed on September 9, 2007 near the West limb. The heliocentric
distance varies between 70$^\circ$ and 75$^\circ$ from left to right.
The granules appear as elongated structures due to the strong
foreshortening in the E-W direction, but otherwise they are clearly
visible. Figure~1{\em b} displays the intensity profiles of the two
\ion{Fe}{1} lines along the slit indicated in panel {\em a}. The slit
crosses several granules and intergranular lanes. Because of the large
heliocentric angles, the typical Doppler shifts associated with
granulation at disk center are no longer seen. Indeed, the spectra
show small wavelength displacements except for the bright granule
marked with horizontal dashes. The intensity profiles emerging from
this granule are remarkable because of their very extended blue
wings. The absorption in the wing is so large that it produces a line
satellite. Similar features have been reported before, but only in
magnetic structures harboring very strong flows as, for example,
sunspot penumbrae (e.g., Bumba 1960; Wiehr 1995; Bellot Rubio
2009). Interestingly, no satellites are observed to the red 
anywhere along the slit.

Figures 1{\em c} and 1{\em d} show intensity profiles from two spatial
positions within the granule. One of them (panel {\em c}) is at the
edge of the granule facing the observer, roughly on the line
connecting it with the disk center (hereafter referred to as the {\em
line of symmetry}). The other position (panel {\em d}) also samples
the outer parts of the granule, but more perpendicularly to the line
of symmetry. The profiles displayed in Figure 1{\em c} exhibit
conspicuous line satellites in the blue wing. The satellite of
\ion{Fe}{1} 630.15~nm has an inflection point that seems to mark the
position of its core. It is located at $\Delta \lambda = -19$~pm from
the center of the average quiet Sun profile at the same heliocentric
angle.  This corresponds to a line-of-sight (LOS) velocity of 9.0
km~s$^{-1}$ towards the observer. Note that the actual flow speed must
be larger due to projection effects.

\begin{figure*}[t]
\centering
\resizebox{.37\hsize}{!}{\includegraphics[bb= 60 360 315 520]{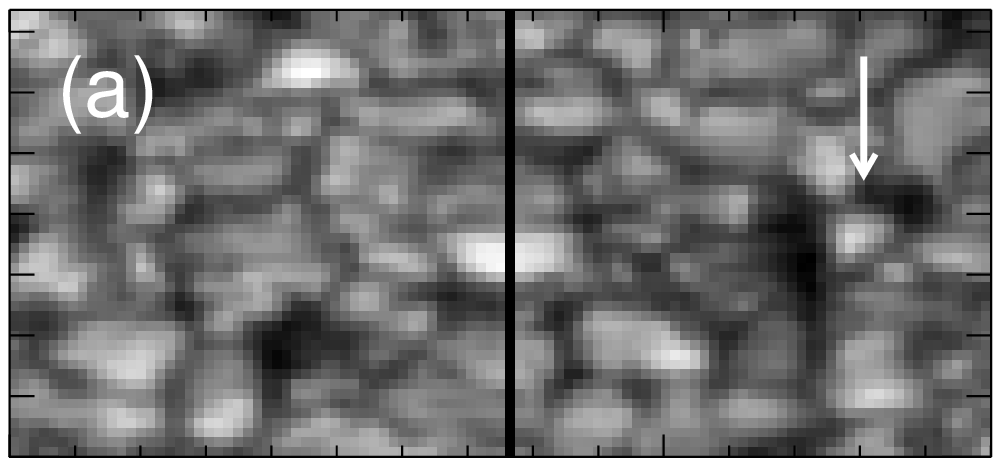}}
\resizebox{.37\hsize}{!}{\includegraphics[bb=38 360 291 520]{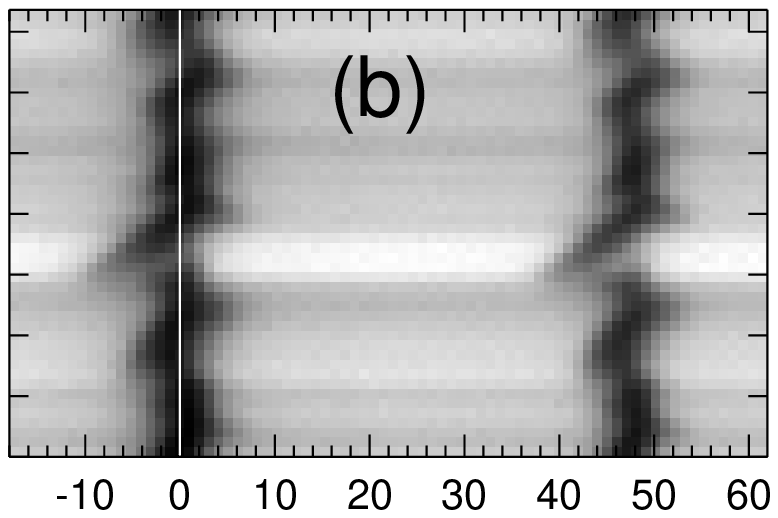} }
\caption{{\em (a)} $15.0\arcsec \times 7.4\arcsec$ region 
scanned with the Hinode/SP on September 24, 2007 at
$\theta=60^\circ$. The arrow indicates the direction to disk
center. Tickmarks are separated by 1\arcsec. {\em (b)} Spectrogram
along the slit marked in panel {\em a}. The wavelength axis is given
in km~s$^{-1}$. The zero velocity (vertical line) corresponds to the 
average line-core position of \ion{Fe}{1} 630.15~nm in the displayed FOV. }
\label{fig2}
\end{figure*}

In principle, absorption at such large distances in the line wing can
also be produced by slower flows, but they would have to be associated
with microturbulent velocities of $\sim 4$~km~s$^{-1}$. Those values
are unrealistically large. Even if they exist, however, LOS velocities
of $\sim 4-5$~km~s$^{-1}$ would still be required to explain the
substantial wavelength shift of the satellites.

The velocities derived from the \ion{Fe}{1} 630.2~nm lines refer to
the optical depth range from $\log \tau_5 \sim 0$ to $-2$ (Ca\-bre\-ra
Solana et al.\ 2005). Using the temperatures, densities, and gas
pressures of the Harvard-Smithsonian Reference Atmosphere (Gingerich
et al.\ 1971), the sound speed can be estimated to be 7.7 and
7.0~km~s$^{-1}$ where the optical depth reaches those values at a
heliocentric angle of 70$^\circ$. The local sound speed may be 
even lower in the cooler granular regions. Thus, the wavelength shifts of
$\sim$9~km~s$^{-1}$ observed in Figure~1{\em c} imply supersonic
flows, with Mach numbers of at least 1.2.

By contrast, the profiles displayed in Figure 1{\em d} do not show
obvious line satellites. They exist, but cannot be clearly identified
because they are blended with the main absorption feature as a
consequence of smaller Doppler shifts. Nevertheless, the flows still
produce strongly distorted profiles with extended blue wings, similar
to those observed by Solanki et al.\ (1996). Solanki et al.\ (1996)
mentioned two different mechanisms capable of producing such broad
profiles: supersonic granular flows, and the shocks associated with
them. In the absence of line satellites they could not favor the 
first scenario, but at the resolution of Hinode it seems clear that 
the distorted profiles are caused by supersonic motions.

Furthermore, the flows producing the satellites and the extended blue
wings have to be essentially radial. This would explain why, in the
external parts of the granule facing the disk center, the wavelength
shift of the satellite is maximum on the line of symmetry and very
small perpendicularly to it. The transition between the two situations
is smooth, as demonstrated by Figure 1{\em b}.

Inspection of Figures 1{\em c} and 1{\em d} reveals that the cores of
the main absorption features are slightly redshifted. The maximum
redshifts occur at the granular edges (where the satellites show
stronger blueshifts), and tend to disappear toward the center of the
granule. The main absorption feature is likely to get a significant
contribution from (or even be entirely produced by) the intergranular
plasma crossed by the slanted lines of sight, which cut the adjacent
intergranular lane before reaching the granule. It is also possible
that granular and intergranular profiles get mixed at the border of
the granule due to the increasingly larger surface area covered by the
pixel far from disk center.  Both scenarios are consistent with the
fact that the line-core redshifts detected at the very edge of the
granule and the intergranular lane next to it are essentially the
same.

The radial variation of the satellites across granules is examined in
Figure 2. The displayed region was located at a heliocentric angle of
60$^\circ$ towards the North solar pole. The vertical line in panel
{\em a} indicates a slit position intersecting a granular cell through
its center. The lower part of the granule faces the observer, while
the upper half is closer to the limb. Figure 2{\em b} depicts the
intensity profiles recorded along the slit. Clearly, the line
satellite shows maximum blueshifts at the border of the granule in the
direction to disk center. The blueshift decreases monotonically as the
limbward edge of the granule is approached. At that position the
satellite has merged with the main absorption feature and is no longer
distinguishable. Beyond the granule, the slit crosses a narrow
intergranular lane and small redshifts are detected. This behavior is
typical and can also be observed in the examples of Figure 5.

The variation of the Doppler shift described above is consistent with
a radial outflow that is vertical at the center of the granule and
more horizontal near the edges. It is likely that the absence of
redshifted satellites on the far side of the granules is due to the
flow occurring below the $\tau=1$ level there. The distance that the
line of sight must travel through the opaque granular plasma before
reaching the horizontal flow is larger in that part of the
granule. This means that the medium could become optically thick well
before the flow is reached; in that case, the flow would be unable to
produce any redshift in the emergent intensity profiles. While the
existence of an opacity effect is clear from the observations, its
origin is still uncertain. A detailed analysis of hydrodynamic
simulations may help understand why the opacity is increased on the
far side of the granules, leading to a complete obscuration of the
flow.

\begin{figure*}[t]
\centering
\resizebox{.95\hsize}{!}{\includegraphics[bb=90 360 1223 1201]{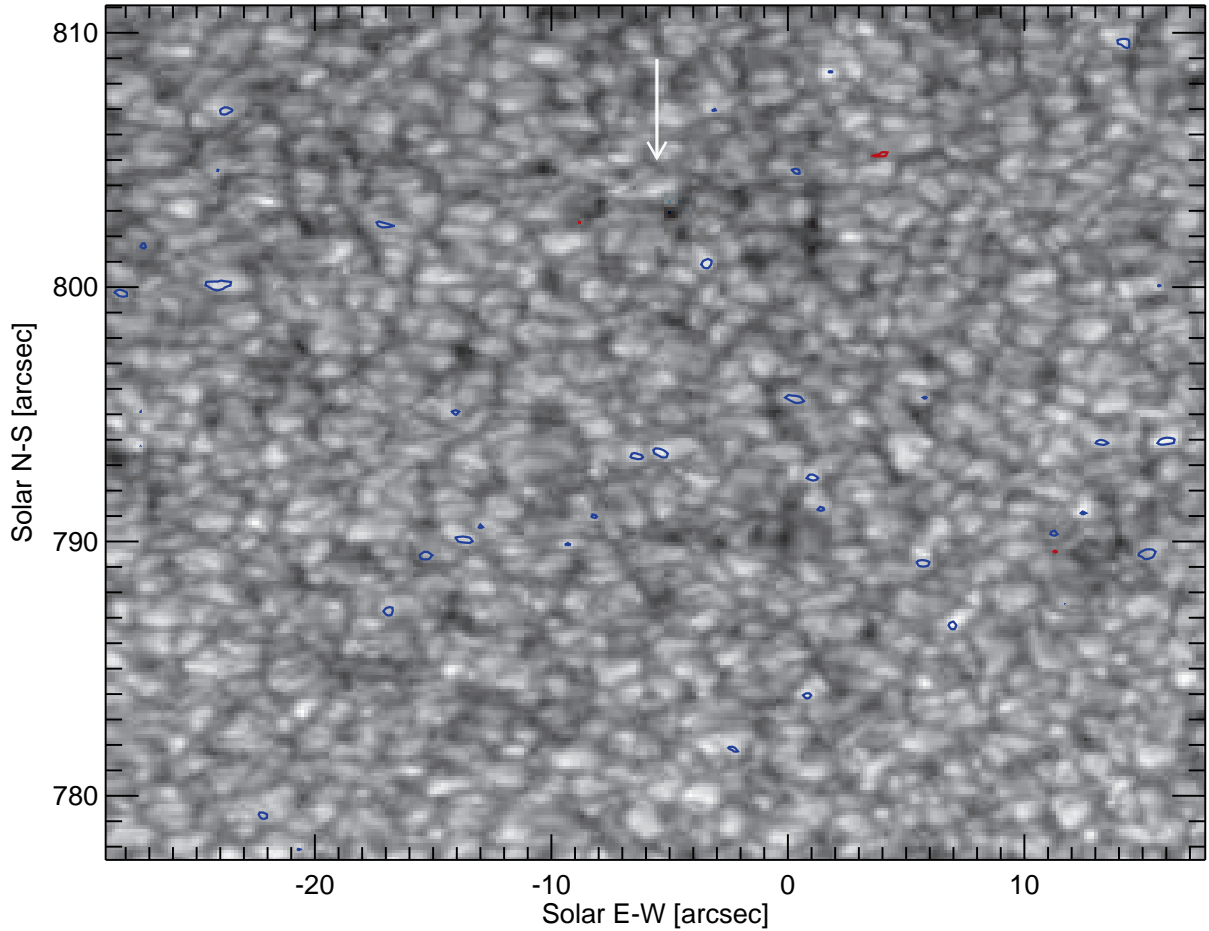}}
\caption{Spatial distribution of supersonic horizontal flows in the solar 
granulation. The displayed region was observed with the Hinode/SP on
September 24, 2007 between 08:15 and 09:24 UT. North is up and West to
the right. The direction to disk center is indicated by the
arrow. Blue and red contours identify locations where the \ion{Fe}{1}
630.15~nm bisectors at the 70\% intensity level are smaller than
$-2.6$~km~$^{-1}$ and larger than $+2.6$~km~$^{-1}$, respectively.}
\label{fig3}
\end{figure*}

As expected for horizontal motions, the line satellites are visible at
relatively large heliocentric angles and not close to disk
center. Their increased visibility towards the limb results from the
progressively larger LOS component of the flow, which induces a
stronger Doppler shift. Very close to the limb, however, the
satellites disappear due to (a) the large surface area covered by
the pixel, which decreases the filling factor of the horizontal flow
and mixes granular and intergranular regions, and (b) the
progressively higher formation height of the line, which eventually
leaves the layers where overshooting convection occurs. In the future,
these effects should be verified using hydrodynamic simulations.

\section{Spatial distribution}
Fast horizontal flows can be identified in the Hinode/SP maps through
a line bisector analysis. The most blueshifted satellites are usually
weak, therefore it is important to use bisectors at high intensity
levels in order to pick out those cases too. After experimenting with
different intensity levels and velocity thresholds, a good proxy for
fast (supersonic) flows is to have bisector velocities of at least
$-2.6$~km~s$^{-1}$ at the 70\% intensity level. For comparison, the
bisector shifts of the 630.15~nm profiles displayed in Figures~1{\em
c} and 1{\em d} are $-2.6$ and $-1.3$~km~s$^{-1}$, respectively. That
is, the first pixel would be selected but the second would not. This
demonstrates that a threshold of $-2.6$~km~s$^{-1}$ is actually very
conservative.


Figure 3 displays the spatial distribution of fast granular flows in a
$46\arcsec \times 34\arcsec$ subfield observed on September 24,
2007. The heliocentric angle varies between approximately 54$^\circ$
to 58$^\circ$ from bottom to top in the map. Blue and red contours
outline positions with bisector velocites smaller than $-
2.6$~km~s$^{-1}$ and larger than $2.6$~km~s$^{-1}$, respectively. As
can be seen, the stronger blueshifts tend to occur on the centerward
side of the granules. Large redshifts are very rare and almost
exclusively restricted to intergranular lanes.

The fraction of pixels with strongly blueshifted bisectors is $\sim
0.3\%$ in the two $46\arcsec \times 164\arcsec$ regions observed with
the Hinode/SP on September 24, 2007 at coordinates $(x,y) \sim
(-5\arcsec, 750\arcsec)$. This value differs from the 3-4\% of surface
area quoted by Stein \& Nordlund (1998) as being covered by supersonic
flows. However, it is important to realize that the two estimates
cannot be compared. The reason is that observations are significantly
affected by projection effects. For example, observational studies do
not detect half of the strong horizontal flows, namely the ones
occurring in the limbward side of the granules. Also, the threshold of
$\pm 2.6$~km~s$^{-1}$ used for the bisector shift at 70\% intensity
level may have excluded many pixels with supersonic flows, simply
because the LOS projection of the velocity vector is smaller than the
adopted threshold. This automatically eliminates pixels away from the
granular edges (since the flow is not completely horizontal there) and
regions close to disk center (because the LOS projection of a
horizontal flow is nearly zero in those regions). Thus, the fractional
area of $\sim 0.3\%$ derived from the Hinode observations represents
only a lower limit.

\begin{figure*}[t]
\centering
\resizebox{.315\hsize}{!}{\includegraphics[bb=130 360 580 770]{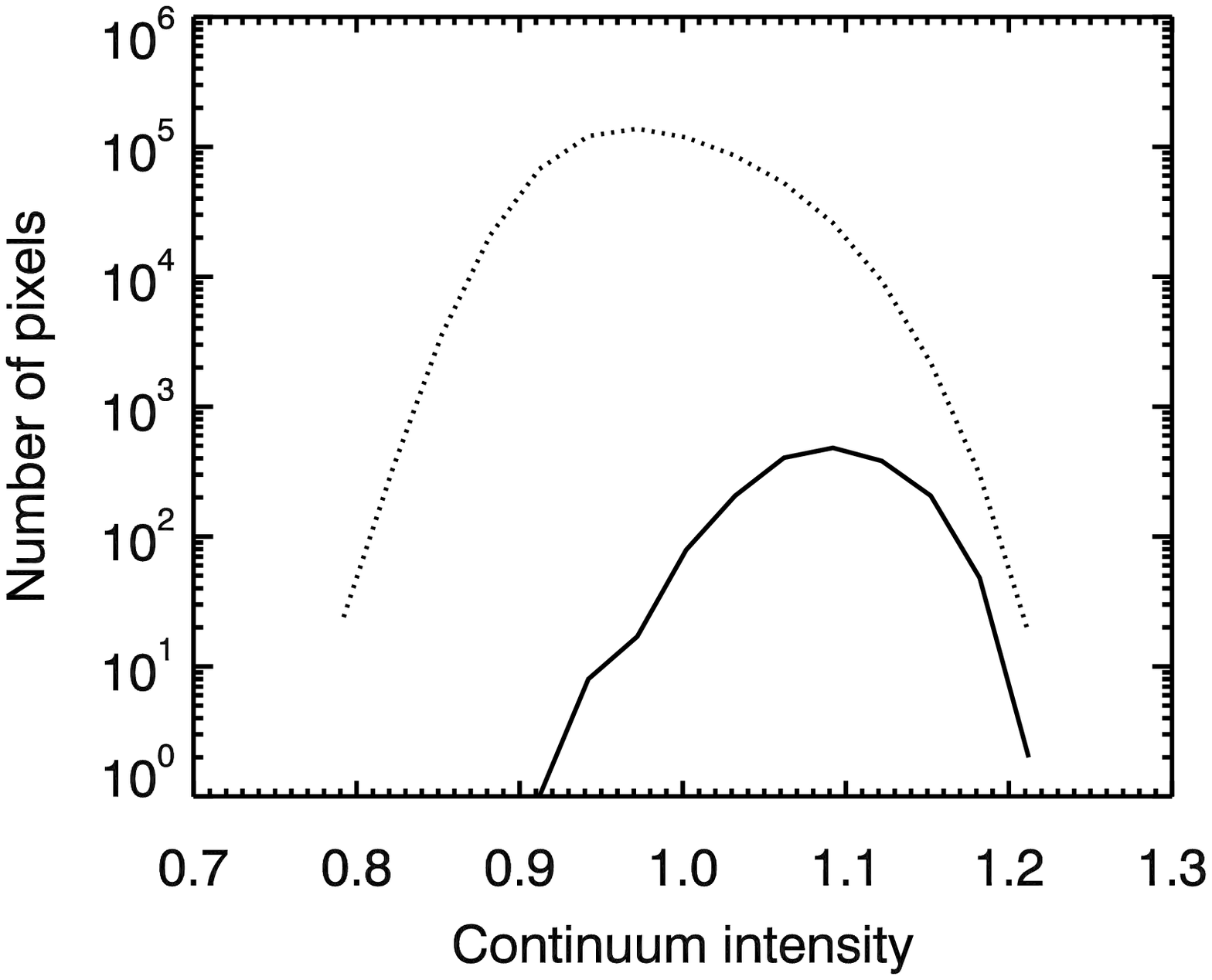}}
\resizebox{.315\hsize}{!}{\includegraphics[bb=130 360 580 770]{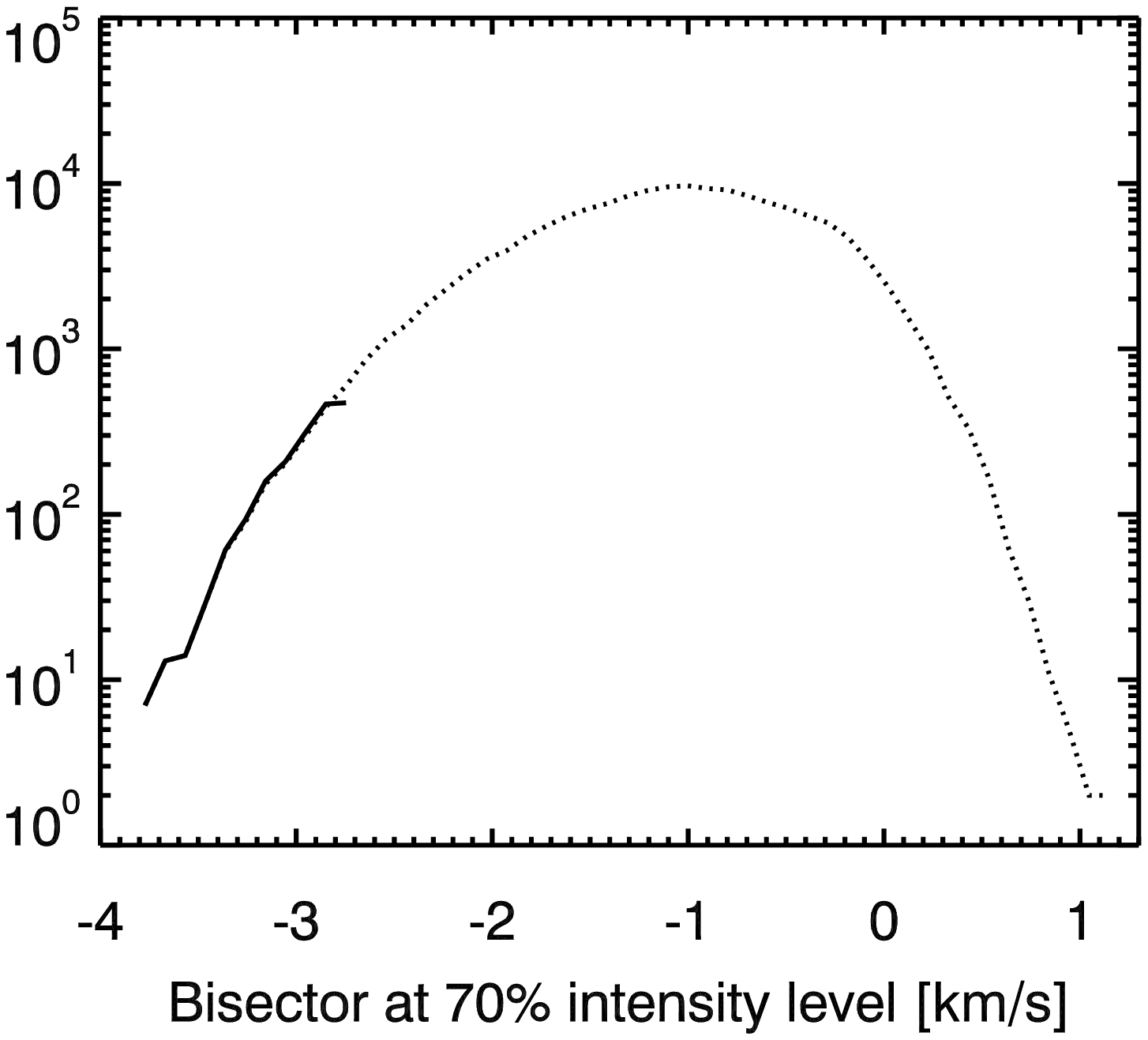}}
\resizebox{.315\hsize}{!}{\includegraphics[bb=130 360 580 770]{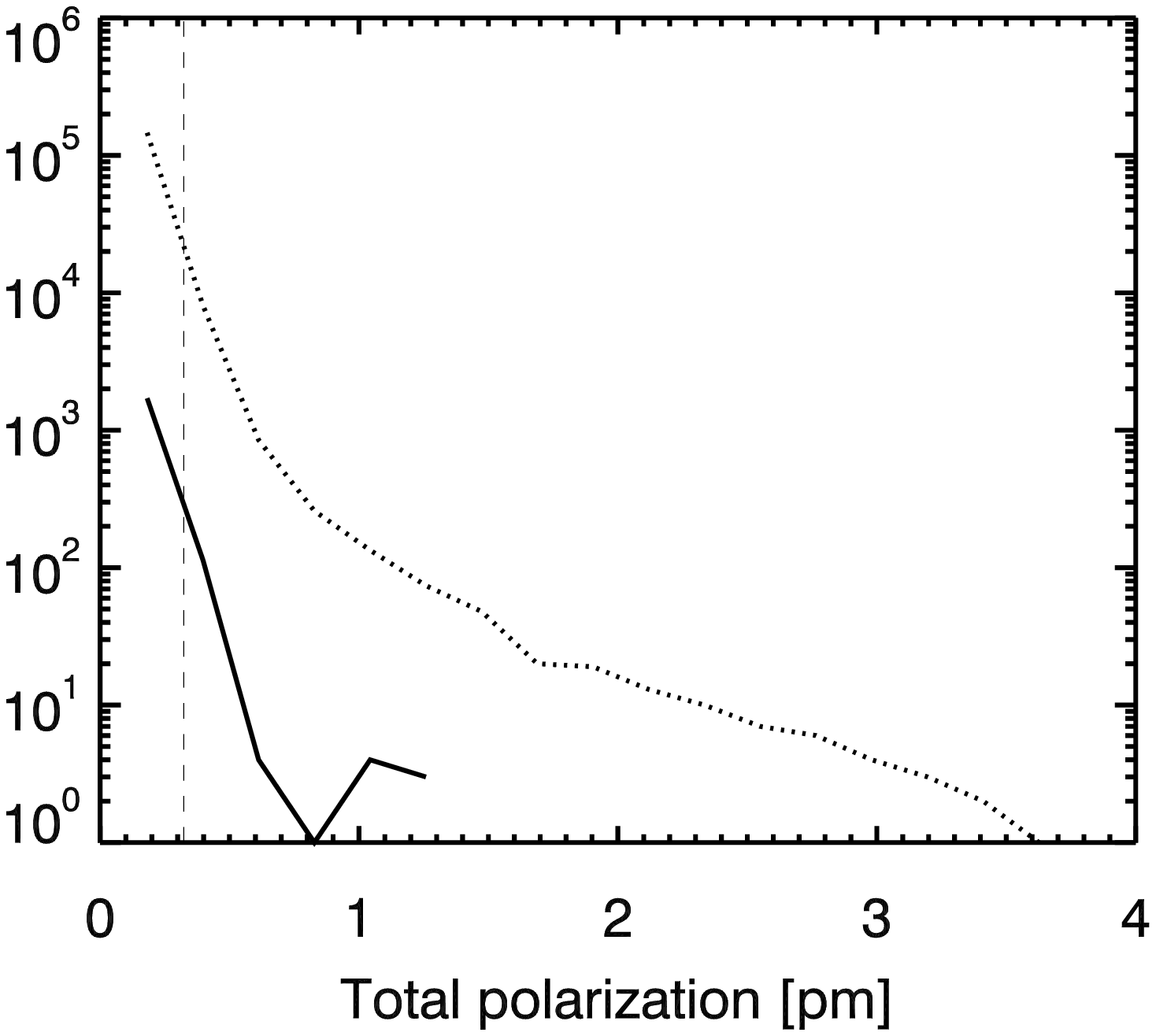}}
\resizebox{.315\hsize}{!}{\includegraphics[bb=130 360 580 770]{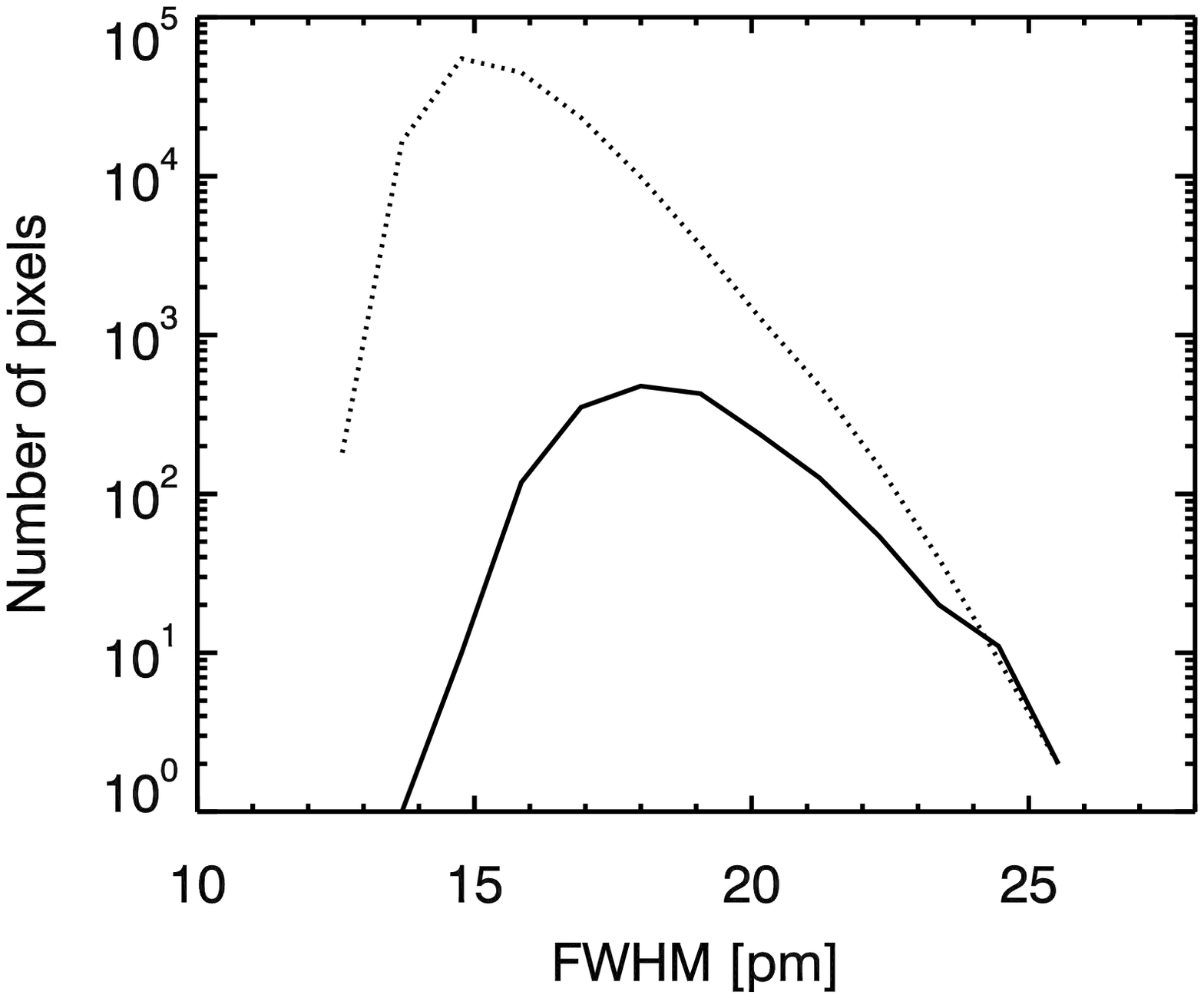}}
\resizebox{.315\hsize}{!}{\includegraphics[bb=130 360 580 770]{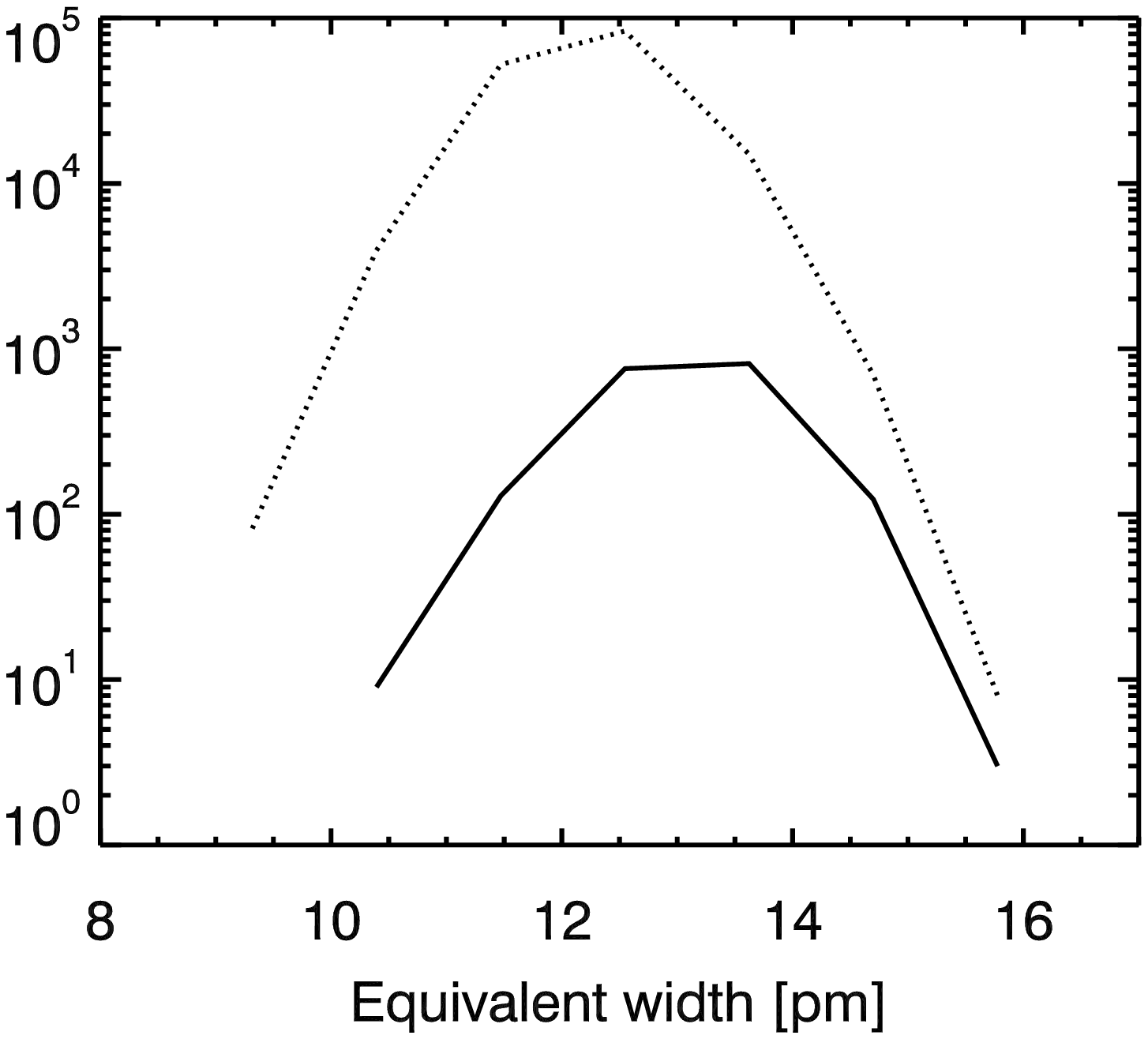}}
\resizebox{.315\hsize}{!}{\includegraphics[bb=130 360 580 770]{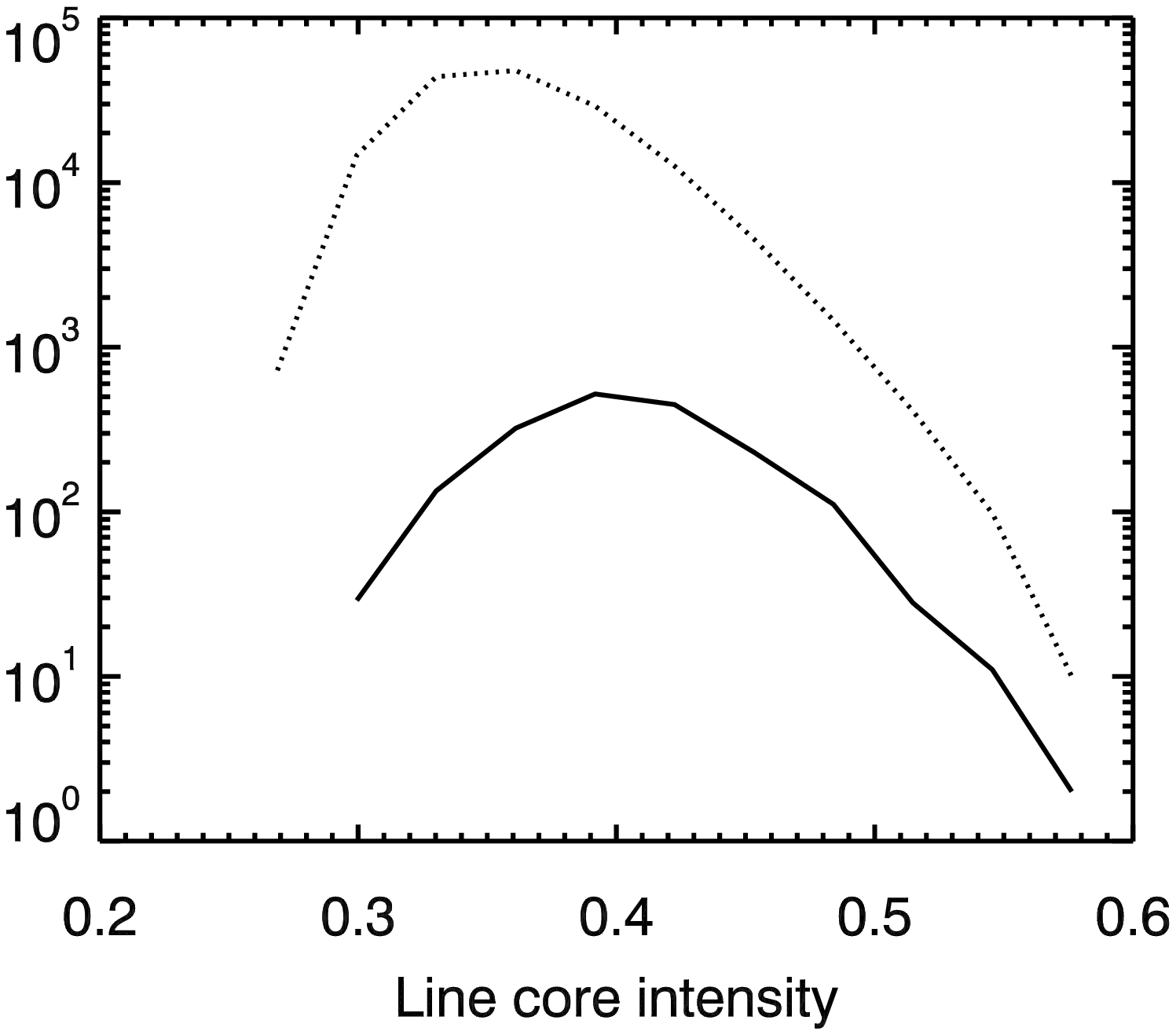}}
\caption{Distributions of line parameters for fast horizontal 
flows (solid) and all granular pixels (dashed) in two scans of 
a quiet region observed with the Hinode/SP on September 24, 2007 
between 08:15 and 09:24 UT, and between 10:10 and 11:19 UT. The
region spans a range of heliocentric angles from $45^\circ$ to
$61^\circ$, and covers a FOV of $46\arcsec \times 164 \arcsec$.
{\em Top:} continuum intensity, bisector velocity at the 70\% intensity
level, and total polarization signal. The vertical dashed line
represents the estimated noise level in total polarization. {\em
Bottom:} Full width at half maximum, equivalent width, and line-core
intensity. All parameters refer to the \ion{Fe}{1} line at
603.15~nm. }
\end{figure*}

\section{Statistical properties}
In this Section the line parameters of pixels harboring fast granular
flows are compared with those observed in granules. As in the previous
Section, fast flows are selected on the basis of their bisector
velocities at the 70\% intensity level. Granules are somewhat
arbitrarily defined as structures with blueshifted line cores and
continuum intensities above 1.02 (in units of the mean continuum
intensity at the same heliocentric distance, to avoid any
center-to-limb variation).

Figure 4 summarizes the results of this analysis for the two regions
observed on September 24, 2007. A total of 1835 pixels with strong
horizontal flows have been identified, while the granular sample
contains 155\,430 pixels (out of 645\,120 pixels in the FOV). Plotted
in the figure are histograms of continuum intensities, bisector
velocities at 70\% intensity level, total polarization signals, full
widths at half maximum (FWHM), equivalent widths (EW), and line-core
intensities, for \ion{Fe}{1} 630.15~nm.

The first thing to note is that both fast horizontal flows (solid
lines) and granules (dashed lines) span a broad range of
parameters. Fast flows tend to be associated with larger values of all
the parameters, except for the total polarization signal. However,
granules without strong horizontal flows may show similarly large
values. This means that detection algorithms based on, e.g., 
enhanced line widths or equivalent widths will produce many false
positives. However, it is also true that the probability of finding
supersonic flows increases significantly at the higher end of the
FWHM, EW, and continuum intensity distributions.  In Figure 4, this
can be seen as a gradually smaller vertical distance between the solid
and dashed lines as the maximum values of the parameters are
approached. An extreme case is provided by the FWHM: all pixels with
FWHM values above $\sim 24$~pm (approximately 1.5 times the average
value in the FOV)
are associated with strong horizontal
motions. Unfortunately, similar statements cannot be made for the
other parameters. Thus, the FWHM is perhaps the best diagnostic of
high-speed flows or shocks in granular convection, as first 
recognized by Solanki et al.\ (1996).

Figure 4 shows that fast granular motions produce bisector velocities
of up to $-3.8$~km~s$^{-1}$ at the 70\% intensity level, although such
extreme Doppler shifts are rare. Another interesting parameter is the
total polarization signal. Hinode has discovered that a large fraction
of granules harbor magnetic fields (Orozco Su\'arez et al.\ 2007). The
distributions of Fig.~4 demonstrate that this is also the case in the
regions analyzed here. However, supersonic flows are not associated
with particularly large polarization signals, as other granular pixels
in the FOV show stronger signals (some of them corresponding to
granules overlaid by the canopies of magnetic flux tubes in facular
and network regions).

\section{Temporal evolution}
It has been suggested, on the basis of numerical simulations, that
supersonic granular motions occur intermittently and exhibit a complex
temporal behavior (Cattaneo et al.\ 1990; Malagoli et al.\ 1990).
These predictions are tested here using time sequences of measurements
near the limb.

Figure~5 shows nine raster scans of a small region located 63$^\circ$
off the disk center. The cadence is 1.9 minutes, a bit long but still
adequate to capture the evolution of granular cells. The upper left
panel displays continuum intensity maps of the region. In the lower
part of the FOV, at $y \sim 1.5\arcsec$, one can distinguish a
relatively large granule that grows, reaches its maximum size, and
then starts to decay. During this sequence of events, line satellites
can be observed near the granular edge facing the disk center, as
illustrated by the spectrograms of Figure~5{\em b}. Bisector
velocities are presented in Figure~5{\em c}. In this and other panels,
contours mark bisector shifts of at least $-2.6$~km~s$^{-1}$.

\begin{figure*}[t]
\centering
\resizebox{.49\hsize}{!}{\includegraphics[bb=65 473 1200 978]{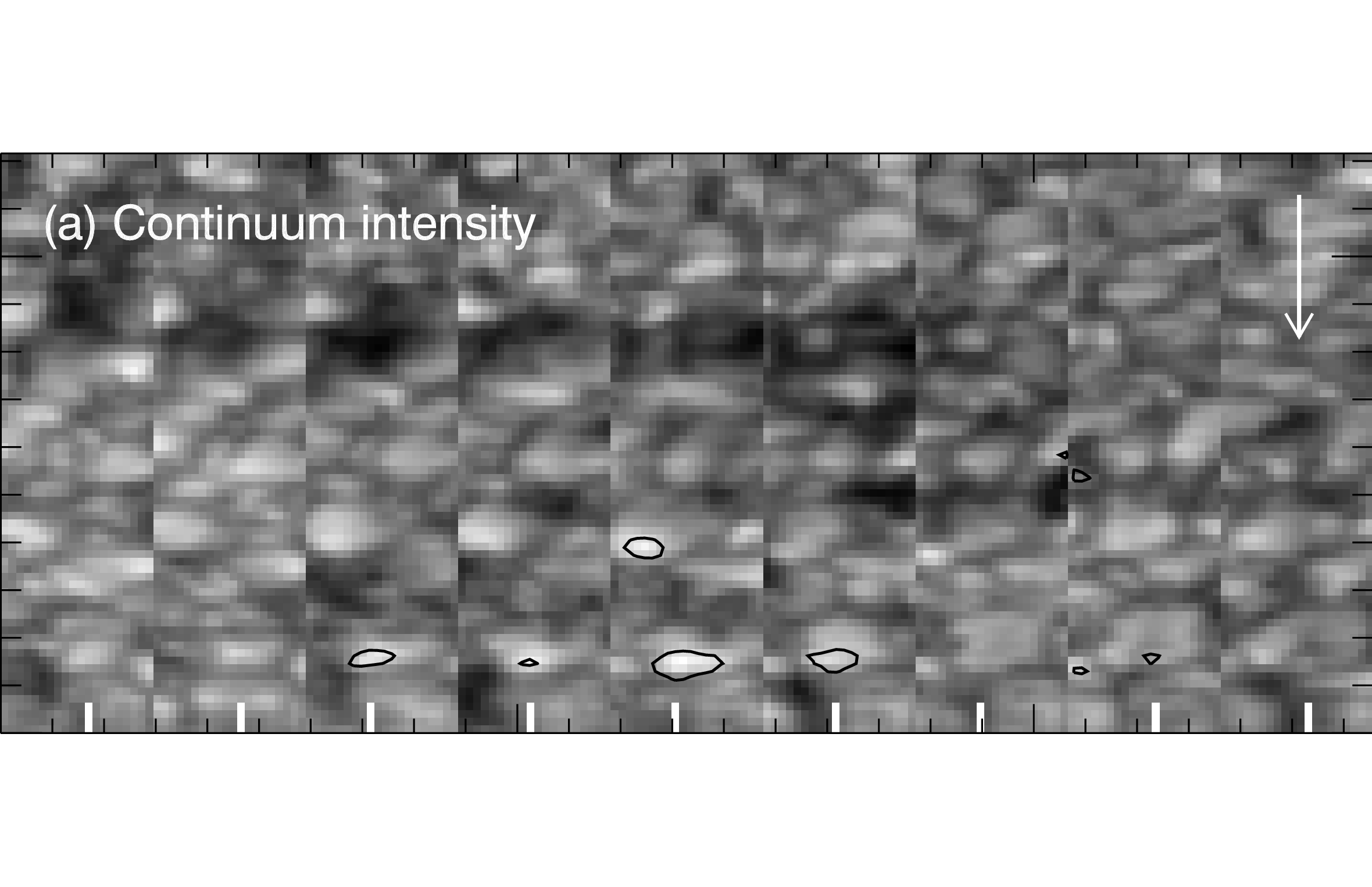}}
\resizebox{.49\hsize}{!}{\includegraphics[bb=44 473 1179 978]{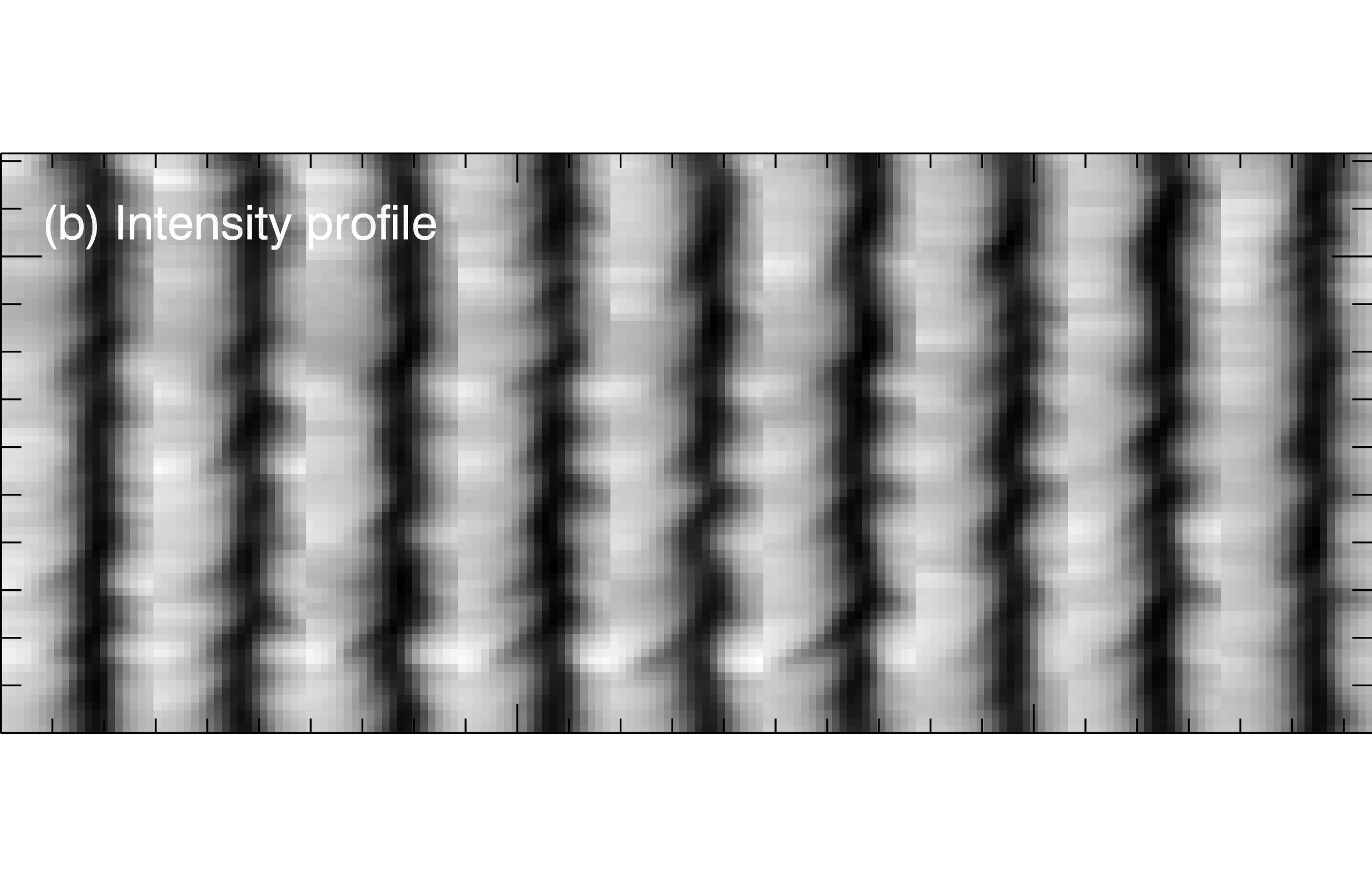}}
\resizebox{.49\hsize}{!}{\includegraphics[bb=65 473 1200 978]{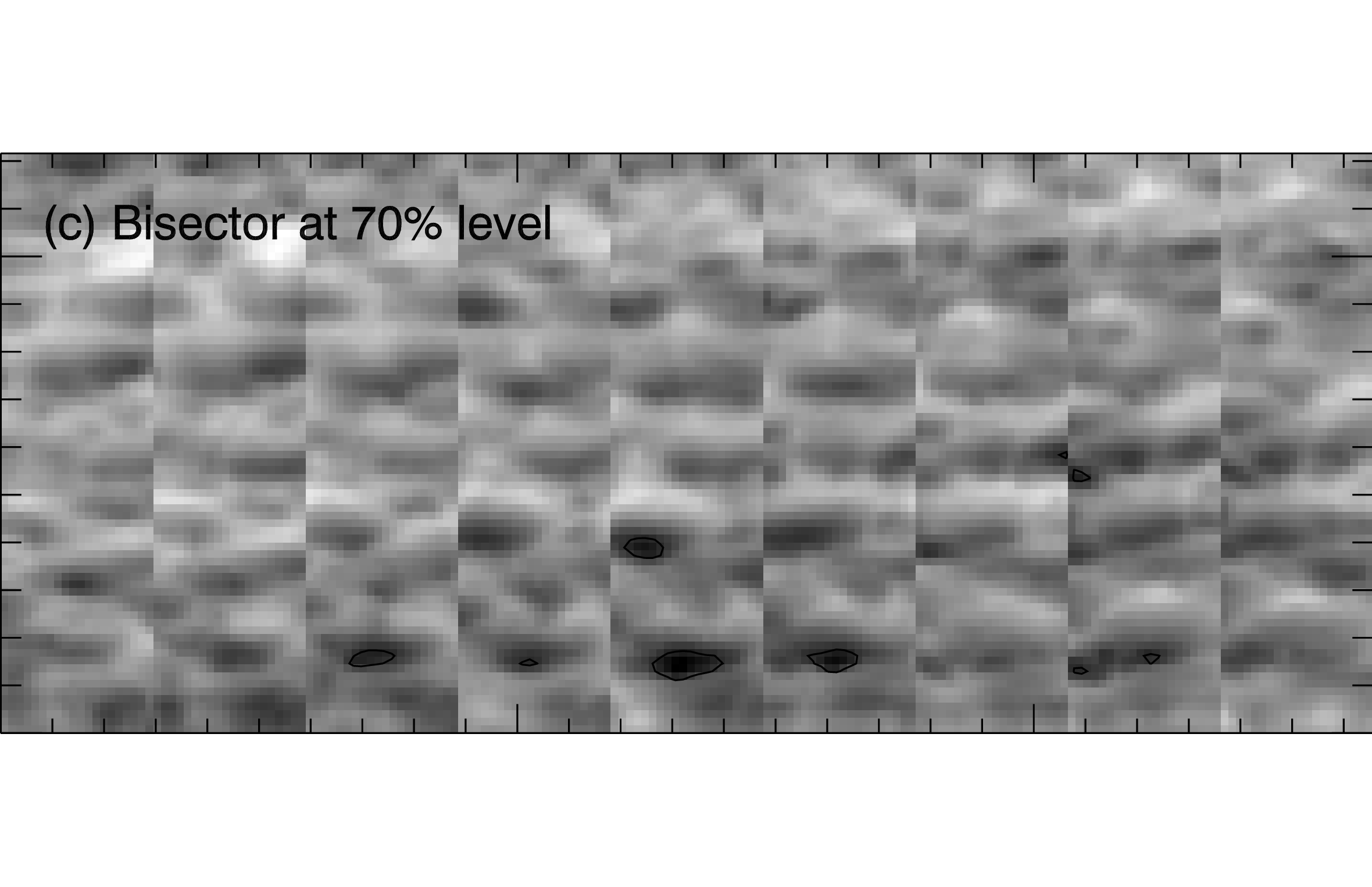}}
\resizebox{.49\hsize}{!}{\includegraphics[bb=44 473 1179 978]{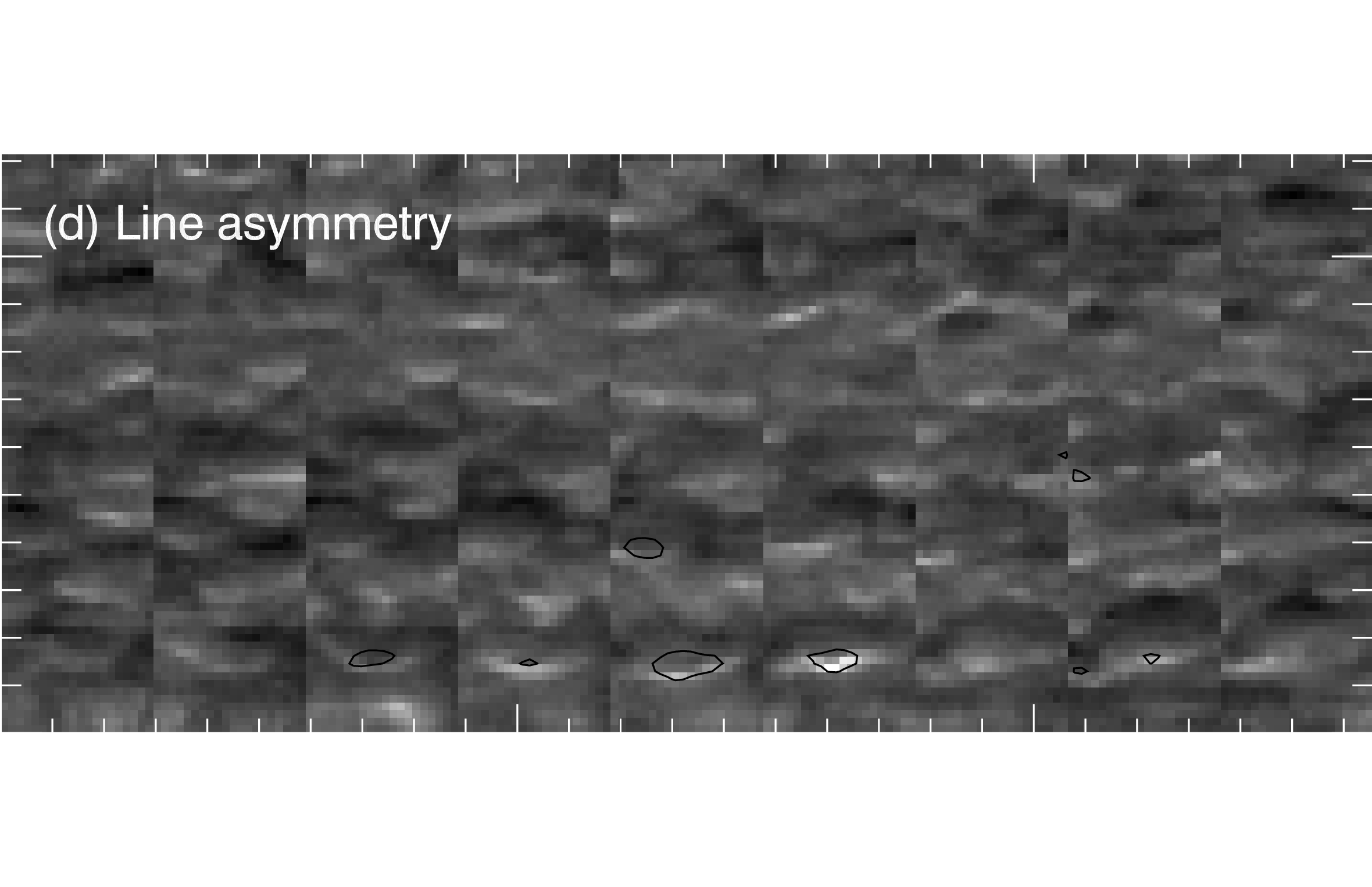}}
\resizebox{.49\hsize}{!}{\includegraphics[bb=65 453 1200 978]{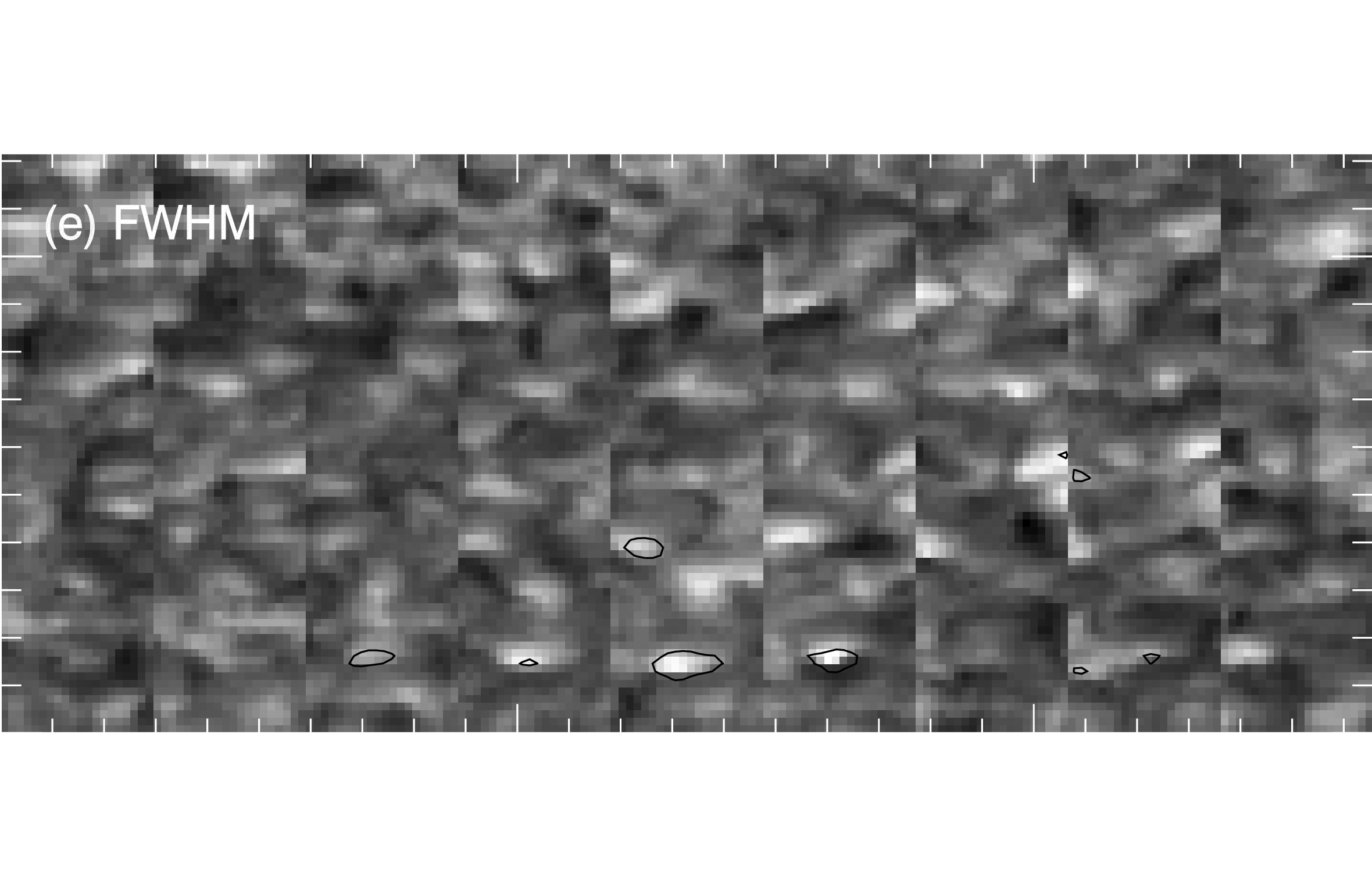}}
\resizebox{.49\hsize}{!}{\includegraphics[bb=44 453 1179 978]{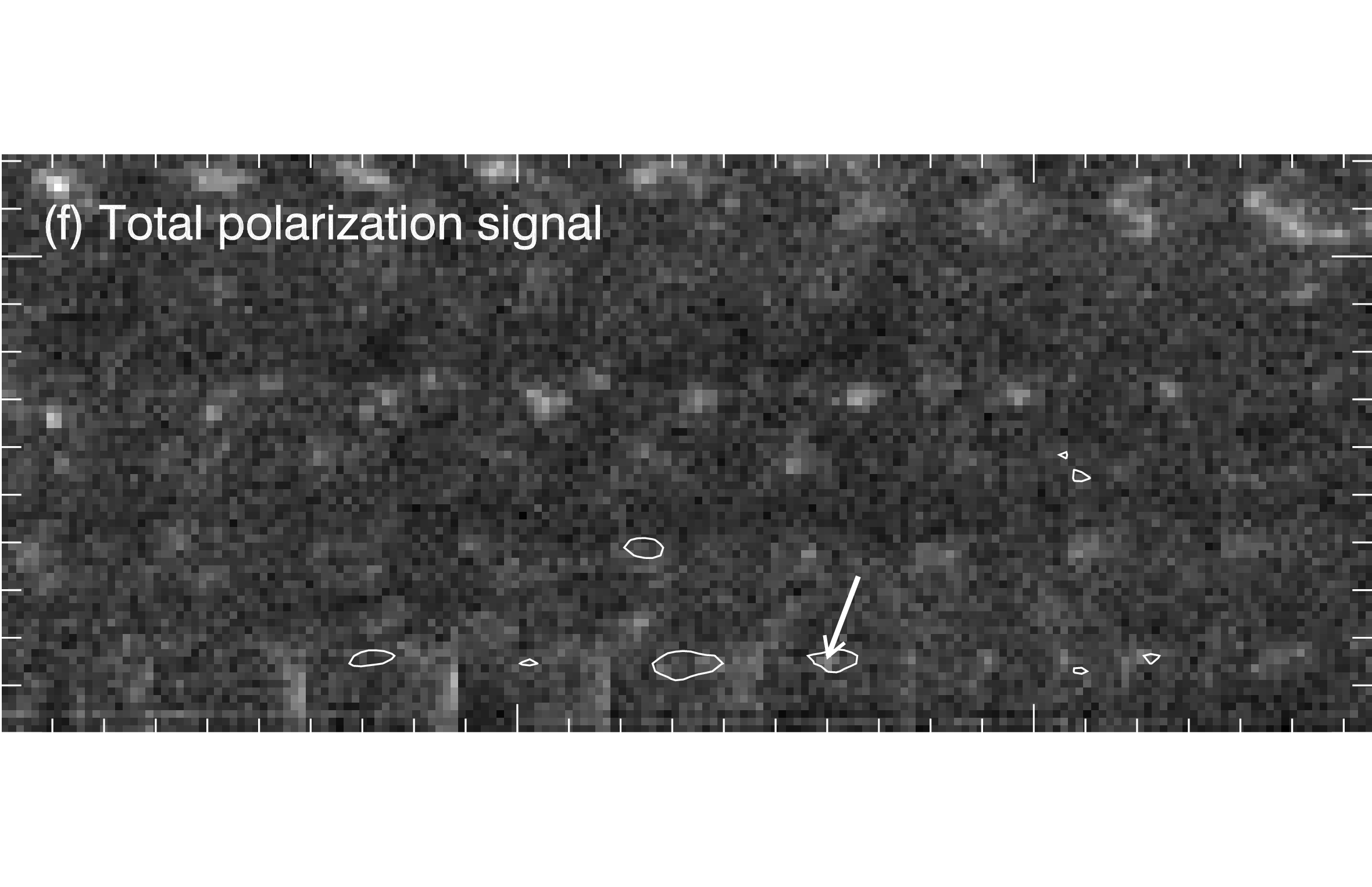}} 
\caption{Temporal evolution of a $3.0\arcsec \times 12.2 \arcsec$ quiet 
Sun area located at a heliocentric angle of 63 deg ($\mu=0.45$). The
observations consist of 9 raster scans made with the Hinode/SP on
September 15, 2007, starting at 19:51:05 UT. The cadence is
115~s. North is up and West to the right. The arrow in panel {\em a}
points to disk center. Tickmarks are separated by 1\arcsec. {\em (a)}
Continuum intensity, scaled between 0.85 and $1.17$ of the average
quiet Sun value. {\em (b)} \ion{Fe}{1} 630.15~nm Stokes $I$ profiles
along the slits marked with white vertical dashes in panel {\em a}. The
wavelength range goes from $-12.4$ to $+7.1$~km~s$^{-1}$. One pixel
corresponds to 2.153~pm, or roughly 1~km~s$^{-1}$. {\em (c)} Bisector
position at 70\% intensity level (from $-3.7$ to
$+2.6$~km~s$^{-1}$). {\em (d)} Degree of line asymmetry in the wing,
measured as the difference of bisector positions at the 50\% and 70\%
intensity levels. The scale ranges from $-0.7$ to
$+2.3$~km~s$^{-1}$. Lage positive values imply lines with very
extended blue wings (line satellites). {\em (e)} FWHM of the intensity
profiles, scaled between 77\% and 137\% of the mean value of
16.2~pm. {\em (f)} Total polarization signal (from 0.22 to
0.55~pm). The arrow shows the location of the Stokes spectra displayed
in Figure 6. In all panels, contours outline structures where 
the bisector at the 70\% intensity level is blueshifted by more than
$2.6$~km~s$^{-1}$ with respect to the average line core position. They
indicate the location of fast horizontal granular flows.}
\label{fig2}
\end{figure*}

During the initial phases of the evolution (first and second scans),
no strong velocities are observed anywhere in the granule. Later, as
the granule grows in size and brightness, blueshifted satellites
appear in its lower half (third to sixth scans). The maximum
blueshifts occur in the fifth scan, about 9.6 minutes after the start
of the time sequence. There is no smooth variation of the LOS velocity
from one map to the next. For example, the bisector shifts are
significantly reduced in the fourth scan, to the extent that the
strong velocities almost disappear. However, 1.9 minutes later the
velocity reaches a maximum. From that point on the blueshifts become
smaller. No bisector velocities faster than $-2.6$~km~s$^{-1}$ are
detected in the seventh scan and only a small patch exhibits strong
plasma motions in the eighth scan.  After that, the granule decreases
in size, fragments, and eventually disappears (not shown). It is
remarkable that the maximum velocities occur while the granule is
still expanding. Once the granule has grown to its full size, the fast
horizontal motions disappear except for the short-lived patch present
in the eighth map. All in all, strong flows are observed only during
less than half of the granular lifetime.

As can be seen in Figure~5{\em e}, the largest blueshifts in the FOV
are associated with enhanced FWHM values. Both occur {\em inside} the
granules close to their borders, not in the neighboring intergranular
space. This agrees with the findings of Nesis et al.\
(1993). Intergranular downdrafts are also the site of enhanced line
broadening (Nesis et al.\ 1992; Solanki et al.\ 1996; Hanslmeier et
al.\ 2008), but their FWHM values are smaller than those detected at
the edges of some granules.

The FWHM enhancements may be generated by supersonic flows decaying
into shock fronts and producing turbulence (Nesis et al.\ 1992). An
alternative explanation is that the FWHM increases because of large
velocity gradients across the shock (Solanki et al.\ 1996; Ryb\'ak et
al.\ 2004). The Hi\-no\-de data favor the second mechanism, as can be
deduced from Figure~5{\em d}. Shown there is the evolution of the line
asymmetry, defined as the difference of bisector shifts at the 50\%
and 70\% intensity levels. Large values of this quantity imply very
extended blue wings and/or line satellites, which can be interpreted
as being due to strong velocity gradients. In Figure~5{\em d}, the
maximum line asymmetry is reached in the sixth scan, coinciding both
in time and space with the maximum of the FWHM.

On the other hand, strong horizontal flows tend to be associated 
with the brightest parts of granules. In the fifth and sixth scans, 
the granule near $y \sim 1.5\arcsec\/$ undergoes a small brightness
enhancement of about $2-3\%$ at the position of the maximum
blueshifts.
The same behavior is observed in the granule at $y \sim 4\arcsec\/$
(fifth scan). Continuum intensity enhancements at the edges of
granules were reported for the first time by de Boer et al.\
(1992). In numerical simulations, the granular edges are brighter 
than the centers because of their larger vertical velocities. The
transient brightenings observed in Figure 5 may be due to stronger
vertical flows or to the development of shocks, which would release
energy and heat the plasma. Distinguishing between the two
possibilities requires an analysis of the evolution of the granular
flow and brightness based on higher cadence observations. 


The two granules of Figure~5 with supersonic flows exhibit weak (but
clear) polarization signals toward the end of their lives, as can be
seen in panel {\em f}. An example of the Stokes spectra emerging from
the edge of the granule at $y \sim 1.5\arcsec\/$ is given in
Figure~6. No linear polarization is detected above the noise level,
hence Stokes $Q$ and $U$ are not plotted. The circular polarization
profiles (Stokes $V$) show three lobes, indicating a complex magnetic
topology at that position. Very remarkably, the Stokes $V$ lobe placed
at shorter wavelengths is strongly blueshifted, just as the line
satellite observed in the intensity profile. This suggests that the
magnetic fields producing the Stokes $V$ signal are associated with
strong horizontal granular motions. One may speculate that these flows
are able to drag magnetic field lines and concentrate them at the
border of the granule or the adjacent intergranular lane, where they
become visible. The field lines would emerge into the photosphere
through the granules and may correspond to the small-scale magnetic
loops discovered by Centeno et al.\ (2007) internetwork regions of
the quiet Sun. Further observations with higher cadences and lower
noise levels are required to confirm or disprove this scenario.

\section{Summary and conclusions}
Hinode observations of the quiet Sun at large heliocentric distances
show that double-peaked intensity profiles occur frequently on the
centerward side of granules. They consist of a main absorption feature
roughly centered at the position of the average profile and a strongly
blueshifted, usually weaker, line satellite. Satellites are never seen
in the red wing of the intensity profiles or on the limbward side of 
the granules.

Sometimes the satellites are nearly resolved, in the sense that their
cores show up prominently away from the main absorption feature. In
those cases, their Doppler shifts indicate supersonic LOS velocities
of up to $\sim 9.0$~km~s$^{-1}$. The sound speed ranges between 7 and
8 km~s$^{-1}$ at the formation level of the \ion{Fe}{1} 630~nm lines,
hence the Mach number of the granular flow is at least 1.2. Even
higher values can be expected in the cooler regions, where the sound
speed may reach 6.5 km~s$^{-1}$.

At the edge of the granular cell, the satellite exhibits the largest
wavelength shift and is usually weak. Toward the granule center, along
the line of symmetry, the satellite becomes stronger but its
wavelength shift decreases, until it eventually merges with (or
replaces) the central absorption feature. This variation of the
Doppler shift is consistent with a radial outflow becoming more
horizontal as the granular edge is approached. The visibility of the
satellites increases toward the limb, due to a more favorable
projection of the velocity to the line of sight. However, very close
to the limb the satellites disappear, presumably because the line is
no longer formed in the layers where overshooting convection
occurs. The exact reason can only be found from a detailed analysis of
hydrodynamic simulations.

\begin{figure}[t]
\centering
\resizebox{.9\hsize}{!}{\includegraphics[bb=47 370 472 793]{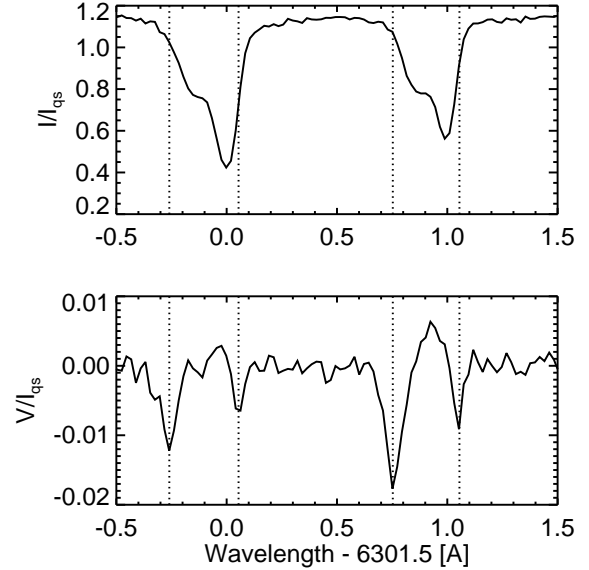}}
\caption{Intensity and circular polarization profiles emerging 
from the edge of a granule harboring strong horizontal flows.  The
exact location of the profiles is indicated with an arrow in the sixth
map of Figure 5{\em f}. Until that moment, no clear polarization
signals were detected in the granule or the adjacent intergranular
lane. The zero of the wavelength scale represents the average line
core position of \ion{Fe}{1} 630.15~nm over the FOV. The vertical 
lines mark the wavelength positions of the outer Stokes $V$ lobes.  }
\label{fig6}
\end{figure}

Highly asymmetric granular spectra were known to exist from earlier
reports (e.g., Solanki et al.\ 1996), but this is the first time that
satellites are detected unambiguously. The discovery has been made
possible by the high spatial resolution, stability, and seeing-free
conditions provided by Hinode. The supersonic flows analyzed in this
paper share some similarities with those observed by Shimizu et al.\
(2008) and Nagata et al.\ (2008) in strong flux concentrations of the
network. The fundamental difference is that the latter are the result
of physical processes involving magnetic fields, while the former are
caused by turbulent convection without the interplay of magnetic
fields.

Supersonic granular flows can be identified through bisector shifts of
more than $-2.6$~km~s$^{-1}$ at the 70\% intensity level. It is
important to go close to the continuum to pick out also the weakest
satellites. In the Hinode scans, at least 0.3\% of the pixels harbor
fast horizontal flows. This value must be regarded as a lower limit
due to strong observational limitations. As such, it is consistent
with numerical models predicting supersonic flows in 3-4\% of the
solar surface at any time (Stein \& Nordlund 1998).

The strongest flows tend to be associated with high continuum
intensities, broad profiles, and large equivalent widths. However, the
correlation is not perfect and high values of these parameters do not
necessarily imply supersonic flows. The Hinode observations suggest
that the enhanced line width (quantified in terms of the FWHM) is due
to large velocity gradients along the line of sight, not to increased
turbulence. The reason is that the largest FWHM values occur {\em within}
the granule where the line asymmetry (defined as the difference of
bisector shifts at the 70\% and 50\% intensity levels) is
maximum. Strong velocity gradients can be expected from shock fronts,
but they could also be the result of rays crossing intergranular lanes
before reaching the granular flow.

As predicted by numerical simulations (e.g., Cattaneo et al.\ 1990;
Malagoli et al.\ 1990), supersonic granular motions are observed to
occur intermittently both in time and in space. First results of an
analysis of time sequences near the solar limb suggest that strong
flows develop only during half or less of the granule
lifetime. However, the cadence of the observations was not optimum and
this result needs to be confirmed with additional observations.

Supersonic flows are expected to produce shocks when they brake
abruptly at the edge of granular cells. Actually, shock signatures
have been observed by Ryb\'ak et al.\ (2004). The data presented here
reveals that the strongest flows are associated with small continuum
intensity enhancements of only 2-3\%. This appears to confirm the
predictions of some numerical simulations that shocks do not cause
significant brightness enhancements (Steffen et al.\ 1994). The
possibility of shocks giving rise to perturbations propagating to
chromospheric layers, however, remains to be investigated.

Supersonic flows could also contribute to the emergence of magnetic
fields in granules. The gentle granular upflows may take field lines
from subsurface layers to the photosphere. Fast horizontal flows would
then concentrate the fields at the granular borders and/or the
neighboring intergranular space, where they would become visible after
reaching a minimum flux density. Support for this scenario is provided
by the observation of weak but clear magnetic signals at the edge of
granules harboring strong horizontal flows. The Stokes $V$ profiles
recorded at those locations are anomalous and often show three
lobes. At least one of them is strongly blueshifted, indicating that
the magnetic field moves with supersonic speeds. It is tempting to
associate this process with the emergence of small-scale magnetic
loops in granules (Centeno et al.\ 2007), but a definite confirmation
requires additional observations with better signal-to-noise ratios.

Given their small spatial extent (less than 0.5\arcsec), a detailed
characterization of high-speed granular flows calls for
nearly-diffraction limited spectroscopic measurements under
seeing-free conditions. At present only Hinode can provide such
observations, but hopefully the Imaging Magnetograph eXperiment (IMaX;
Mart\'{\i}nez Pillet et al.\ 2004) aboard SUNRISE (Barthol et al.\
2006) will be able to take them in the near future. IMaX will not only
improve the spatial resolution up to 0.1\arcsec (which is especially
important for limb observations), but will also permit faster scans 
due to the large photon collecting power of its 1m telescope.

\acknowledgments 

Hinode is a Japanese mission developed and launched by ISAS/JAXA, with
NAOJ as domestic partner and NASA and STFC (UK) as international
partners. It is operated by these agencies in co-operation with ESA
and NSC (Norway). This work has been partially funded by the Spanish
MICIN through projects ESP2006-13030-C06-02 and PCI2006-A7-0624, and 
by Junta de Andaluc\'{\i}a through project P07-TEP-2687.

\end{document}